\documentclass{jfm}
\usepackage{graphicx}
\usepackage{epstopdf, epsfig}
\usepackage{xcolor}
\usepackage{amsfonts}
\newcommand{\bigepsilon}{\makebox{\large\ensuremath{\epsilon}}}

\usepackage{upgreek}

\usepackage{relsize}
\usepackage{bm}
\usepackage{amsmath}
\usepackage{cleveref}
\crefformat{section}{\S#2#1#3}
\crefformat{subsection}{\S#2#1#3}
\crefformat{subsubsection}{\S#2#1#3}
\crefrangeformat{section}{\S\S#3#1#4 to~#5#2#6}
\crefmultiformat{section}{\S\S#2#1#3}{ and~#2#1#3}{, #2#1#3}{ and~#2#1#3}

\graphicspath{{figures/}}

\shorttitle{Large-scale circulation of turbulent Rayleigh-B\'{e}nard convection}
\shortauthor{V. T. Vishnu, A. K. De and P. K. Mishra}
\title{Reorientation of large-scale circulation of turbulent Rayleigh-B\'{e}nard convection in a cubic cell}

\author{V. T. Vishnu\aff{1},
  A. K. De\aff{1}
  \corresp{\email{akd@iitg.ac.in}},
 \and P. K.  Mishra\aff{2}}

\affiliation{\aff{1}Department of Mechanical Engineering, Indian Institute of Technology, Guwahati 781039, India 
\aff{2}Department of Physics,  Indian Institute of Technology, Guwahati 781039, India}

\begin{document}

\maketitle

\begin{abstract}
We present a direct numerical simulation on the dynamics of large-scale circulation (LSC) of turbulent Rayleigh-B\'{e}nard convection of air ($\Pran=0.7$) contained in a cubic cell for Rayleigh number range $2\times 10^6\le Ra\le 10^9$. The strength and orientation of LSC is quantified by the amplitude ($A_1$) and phase ($\Phi_1$) of the first Fourier mode of the vertical velocity. 
The plane containing LSC is generally aligned along one of the diagonals of the box accompanied by a four-roll structure in the other.
However, an abnormal single-roll state with substructures is noted at the face planes for low $Ra$.
In addition to the primary roll, two secondary corner-roll structures are also observed in the LSC plane which grow in size and destabilise the LSC resulting in partial ($\Delta \Phi_1 \approx \upi/2$) and complete ($\Delta \Phi_1 \approx \upi$) reversals. 
In addition to previously reported rotation-led reorientations, we also observe cessation events which are rare in cubic cells.
The distribution of turbulent kinetic energy shows that the energy is primarily extracted from the non-LSC plane and fed to the LSC plane by the transport mechanism.
We observe that the corner-rolls reduce in size and the substructures diminish at higher $Ra$, which leads to the reduction in the occurrence of reorientations of LSC.

\end{abstract}


\section{Introduction}\label{sec:Intro}
\par
Reorientation of large-scale flow, such as wind reversals in atmospheric flows and the change in polarity of magnetic field in Earth and Sun have been an interesting field of research in turbulence since the last couple of decades. Understanding the genesis of large-scale circulation (LSC) and the mechanism that leads to change in its direction remain a challenge for researchers \citep{Ahlers_Grossmann_Lohse}.
Rayleigh-B\'{e}nard convection (RBC), in which a fluid layer is heated from below and cooled from top, has become a paradigm to address the questions pertaining to LSC and its dynamics.
It has been observed that the vertical plane containing LSC changes its orientation in the azimuthal/lateral direction swiftly and randomly. Such sudden and significant change in the orientation of LSC, known as ``reorientation'', has been observed to occur either as ``rotation-led" in which reorientation takes place without significant change in the strength of the LSC or ``cessation-led" where the measured strength of LSC vanishes during the reorientation and the flow resumes in an arbitrary direction \citep{Brown_Ahlers_JFM,MishraDe}. 
The dynamics of the mean wind is generally investigated in the cylindrical geometry owing to its azimuthal symmetry which facilitates the freedom of orientation of LSC along any plane. However, the absence of such symmetry in box imposes restrictions over the appearance of LSC as well as its dynamics.
In this paper, we focus on the flow structure and mechanism of reorientation of LSC in a cubic cell for highly turbulent flow.
\par
A number of numerical and experimental studies in 2D/quasi-2D systems have shown that in addition to LSC there exists two secondary corner rolls \citep{Sugiyama_Q2D_2010,chandra_Verma_2013,Wagner_Shishkina_2013}. 
With time, interaction among the corner-rolls owing to the growth in their size and kinetic energy leads to flow reversals.
Recently, \citet{Xin_etal_JFMR_2019} studied the emergence of substructures inside the LSC which induces flow reversals in quasi-2D systems.
It has been established that corner-rolls do exist in 3D flows \citep{Sun_etal_2005a,Foroozani_etal_2017}, though unlike its 2D counterpart they do not remain strictly confined in the LSC plane. 
On the other hand, \citet{Valencia_etal_IJHMT_2007} reported that LSC is aligned along the diagonals of the box and it rotates about the vertical axis. 
Diagonal switching \citep{Vasiliev_etal_IJHMT_2016} of this coherent structure leads to flow reorientations \citep{Liu_Ecke_2009,Bai_etal_PRE_2016}.
Recent study of \citet{Foroozani_etal_2017} reported these reorientations as only ``rotation-led" with no signs of ``cessation-led" reversals where angular change varied as widely as $45^{\circ} \leq \Delta \Phi \leq 135^{\circ}$. 
On a different note, \citet{Vasiliev_etal_2018} argued that diagonal rolls are superposition of a pair of large-scale rolls aligned along the faces and reversal of them reflects in change of orientation in the diagonals. \citet{Giannakis_etal_Koopman_JFM_2018} predicted four stable states of LSC, along the two diagonals with clockwise and anti-clockwise sense of rotation, by projecting the velocity field on the Koopman eigenfunctions.
 
\par
Although flow reversals have been extensively studied over the years in 2D/quasi-2D systems and cylindrical domains, origin and dynamics of such large-scale structures in a Cartesian box are not well understood, especially for highly turbulent flow. In the present study, we examine the dynamics of LSC with prime focus on its orientation and subsequent reversals for high Rayleigh number range $2\times10^6\leq Ra \leq 10^9$.
Based on the global energy transport between the mean flow and turbulent background  we quantify a strong correlation between the LSC and turbulent noise which is further connected with the reorientation.
In \cref{sec:numerical} we present the details of the mathematical formulation and numerical details. In \cref{sec:results} first identification of LSC is introduced followed by characterisation of its dynamics which is then substantiated by energy transfer analysis. Finally all the major findings are summarised in \cref{sec:Conclusion}.

\section{Numerical details}\label{sec:numerical}
The governing equations of RBC under  Boussinesq approximation are
\begin{equation}
\frac{\partial u_i}{\partial x_i}=0,
\label{gov_eqn1}
\end{equation}
\begin{equation}
\frac{\partial u_i}{\partial t} + \frac{\partial (u_i u_j)}{\partial x_j}= -\frac{\partial p}{\partial x_i} +
\sqrt{\frac{Pr}{Ra}}\frac{\partial^2 u_i}{\partial x_j \partial x_j} + \delta_{iy}\theta,
\label{gov_eqn2}
\end{equation}
\begin{equation}
\frac{\partial \theta}{\partial t} + \frac{\partial (u_j \theta)}{\partial x_j}= 
\sqrt{\frac{1}{RaPr}}\frac{\partial^2 \theta}{\partial x_j \partial x_j}, 
\label{gov_eqn3}
\end{equation}
where $u_i(u, v,w)$ represents the velocity components in $x,~y$ (buoyancy direction), and $z$ directions, $p$ the pressure, and $\theta$ the non-dimensional temperature, while  $Ra$ and $Pr$ are the Rayleigh number and Prandtl number, respectively.

\par
The above coupled equations (\ref{gov_eqn1})-(\ref{gov_eqn3}) are solved using a predictor-corrector based finite volume formulation with collocated arrangement of variables. No-slip velocity condition is implemented on all the boundaries, while for temperature isothermal and adiabatic conditions are applied on the horizontal and vertical walls, respectively. The non-linear convective term is approximated using the $2$nd-order Adams-Bashforth scheme, while the buoyancy, pressure and diffusive terms are approximated by the $2$nd-order Crank-Nicolson scheme. In order to resolve the boundary layers adequately refined meshes are used near the walls that ensures $(N_{BL})$ 6-10 grid points inside them, while global maximum $(\Delta_{max}$) spacing is kept mostly below the theoretical estimate of the Kolmogorov length scale $\eta$ ($=Pr^{1/2}Ra^{-1/4}Nu^{-1/4}$). Time increment $\Delta t$ ($5\times10^{-4}$ for $Ra \geq 5 \times10^7$ and $10^{-3}$ for the rest) is appropriately chosen to limit the Courant number to $0.2$ \citep{Vishnu_PoF_2019}. All the resulting sparse linear systems are solved using BiCGSTAB technique preconditioned by a highly scalable diagonalised version of Stone's Strongly Implicit Procedure. Global Nusselt number computed by the mean heat flux at the horizontal surfaces: $Nu_S = \langle \partial \theta /\partial y \rangle_{A,t} $, thermal dissipation rate: $Nu_{\bigepsilon_\theta}={(RaPr)}^{1/2}\langle \bigepsilon_\theta\rangle$, and viscous dissipation rate: $Nu_{\bigepsilon_u}={(RaPr)}^{1/2}\langle \bigepsilon_u \rangle+1$ where $\langle ..  \rangle$ and $\langle .. \rangle_{A,t}$ represent volume-time and horizontal area-time averaging are shown in table \ref{tab:Grid} with other numerical details. All the values are in good agreement and they compare well with $Nu_{ref}$ reported by \citet{Kaczorowski_Xia_2013} which indicates satisfactory resolution of the present data.
\begin{table}
\begin{center}
\begin{tabular}{lcccccccc}
$Ra$   & $N_x\times N_y\times N_z$ & $N_{BL}$& $Nu_{S}$ &$Nu_{\bigepsilon_\theta}$&$Nu_{\bigepsilon_u}$ &$Nu_{ref}$&$\Delta_{max}/\eta$ \\ [3pt]
$2\times 10^6$& $100\times100\times100$ &$10$&$10.1$&$10.1$&$10.2$&$10.0$&$0.954$\\
$3\times 10^6$& $100\times100\times100$ &$7$&$11.5$&$11.3$&$11.4$&$11.4$ &$0.963$\\
$5\times 10^6$& $168\times168\times168$ &$13$&$13.2$&$13.2$&$13.4$&$13.4$&$0.927$\\
$1\times 10^7$& $168\times168\times168$ &$11$&$16.0$&$16.0$&$16.2$&$16.2$&$0.962$\\
$2\times 10^7$& $200\times200\times200$ &$8$&$19.9$&$19.6$&$19.8$&$19.6$ &$0.990$\\
$3\times 10^7$& $200\times200\times200$ &$7$&$22.3$&$21.3$&$22.0$&$22.0$ &$0.985$\\
$5\times 10^7$& $264\times264\times264$ &$8$&$25.7$&$25.3$&$25.4$&$--$   &$0.964$\\
$1\times 10^8$& $300\times300\times300$ &$8$&$31.1$&$30.8$&$30.8$&$31.4$ &$0.964$\\
$2\times 10^8$& $372\times372\times372$ &$7$&$38.2$&$37.6$&$37.8$&$--$   &$0.989$\\
$5\times 10^8$& $420\times420\times420$ &$6$&$50.7$&$49.3$&$49.3$&$--$   &$1.157$\\
$1\times 10^9$& $512\times512\times512$ &$6$&$62.1$&$60.6$&$60.6$&$--$   &$1.188$\\
\end{tabular}
\caption{Columns from left to right indicate, $Ra$, total number of grid points ($N_x \times N_y \times N_z$), number of grid points in the thermal boundary layer ($N_{BL}$), Nusselt number computed by surface gradient and dissipation rates ($Nu_S, Nu_{\bigepsilon_\theta}, Nu_{\bigepsilon_u}$), Nusselt number ($Nu_{ref}$) reported by \citet{Kaczorowski_Xia_2013}, and the maximum grid spacing normalised by the Kolmogorov length scale $\eta =Pr^{1/2}Ra^{-1/4}Nu^{-1/4}$.}
\label{tab:Grid}
\end{center}
\end{table}

\section{Results}\label{sec:results}
In this section, we discuss the principal findings of the present work. 
At the outset, the dynamics of LSC and the mechanism of flow reversals are presented. This is followed by analysis of the energy transport in connection with the flow reversals.

 \subsection{LSC and associated dynamics}
\begin{figure}
\centerline{\includegraphics[width=4.0cm]{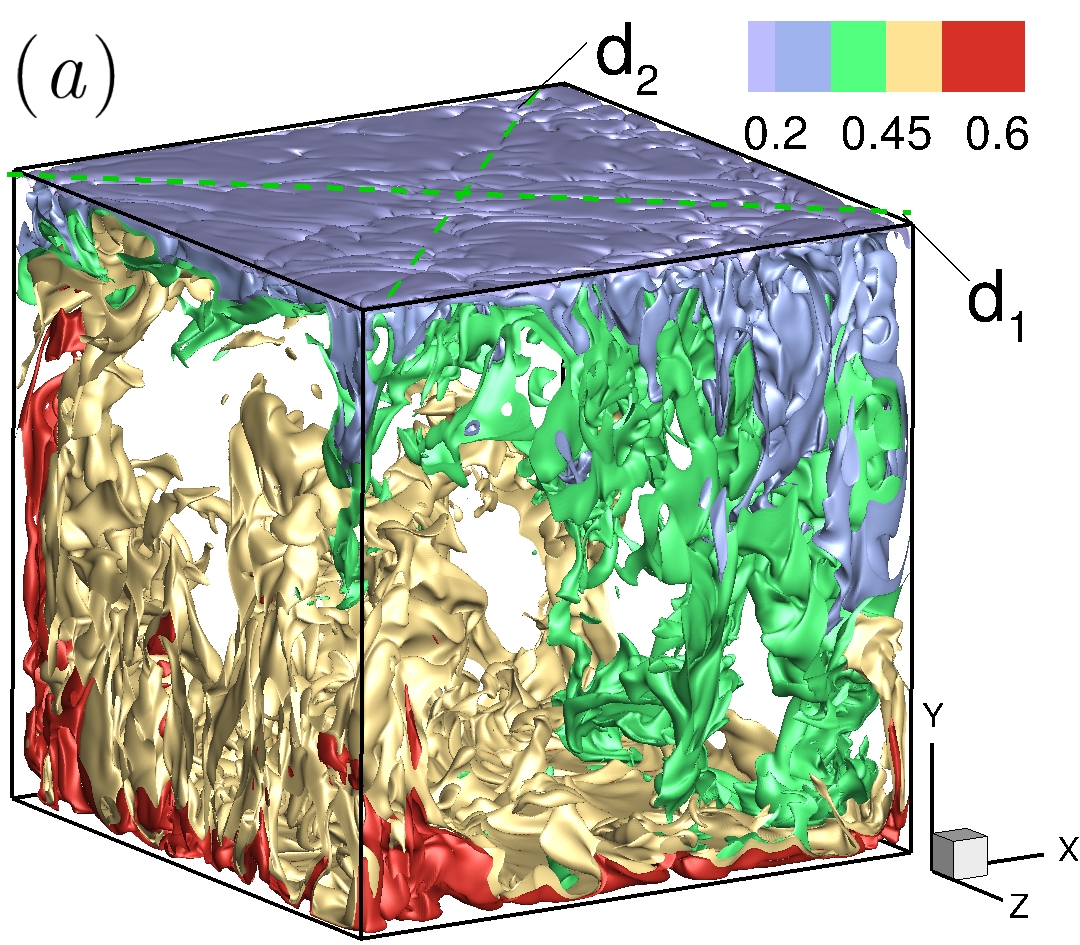}\includegraphics[width=8cm]{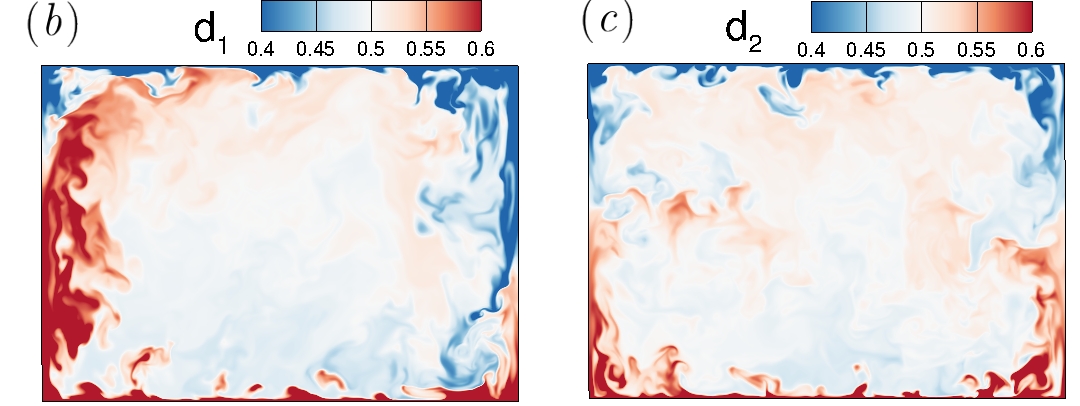}}
\caption{ (\textit{a}) Instantaneous temperature iso-surfaces indicating LSC oriented along the diagonal plane $\mathrm{d_1}$ for $Ra=10^9$. The contours of temperature in the diagonal planes (\textit{b}) $\mathrm{d_1}$ and (\textit{c}) $\mathrm{d_2}$.} 
\label{fig:Flowstr}
\end{figure}

\begin{figure}
\centerline{\includegraphics[width=13.cm]{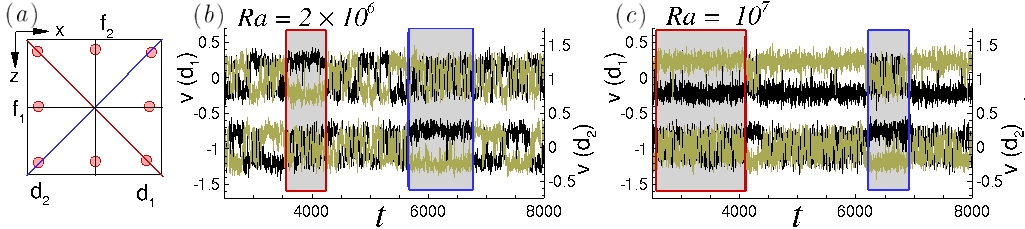}}
\caption{ (\textit{a}) Schematic diagram of the mid-plane ($y=0.5H$) showing the location of numerical probes along with the diagonal and face planes. Time signals of vertical velocity from numerical probes placed across the two diagonals for (\textit{b}) $Ra=2\times10^6$ and (\textit{c}) $Ra=10^7$.The red and blue windows indicate alignment of LSC along the diagonal planes $\mathrm{d_1}$ and $\mathrm{d_2}$, respectively.} 
\label{fig:timesig}
\end{figure}

A sample view of LSC is shown in figure \ref{fig:Flowstr}(\textit{a,b}) for $Ra=10^9$ where hot plumes erupt from the boundary layer and cold plumes sink along the opposite corners giving rise to a large-scale coherent structure aligned along the diagonal labeled as $\mathrm{d_1}$.
However, such a coherent structure is absent in the other diagonal (figure \ref{fig:Flowstr}\textit{c}).
In order to identify and characterise the LSC, time traces of vertical velocity are sampled at eight azimuthally equispaced locations, shown in figure \ref{fig:timesig}(\textit{a}), at the mid-vertical plane located $0.1H$ distance from the lateral walls. The time signals from the probes located across the two diagonals are shown in figure \ref{fig:timesig}(\textit{b,c}), where red and blue windows indicate alignment of LSC along the diagonal planes $\mathrm{d_1}$ and $\mathrm{d_2}$, respectively. 
Along the diagonal plane containing LSC the velocity signals fluctuate about a finite mean value, while in the opposite diagonal they exhibit high variance fluctuations with zero mean indicating the absence of any coherent structures.
With time the flow structures interchange between the diagonal planes suggesting the reorientation of LSC.
We observe that LSC remains confined in a particular diagonal plane for longer duration for higher $Ra$ as evident from figure \ref{fig:timesig}(\textit{b,c}).

Following the works of \citet{MishraDe}, to get more insights into the dynamics of LSC, we carry out the Fourier mode analysis of vertical velocity measured at $N(=8)$ probes in the mid-plane (see figure \ref{fig:timesig}\textit{a}). The Fourier mode is given by
\begin{eqnarray}
 \hat{u}_{k}= \sum_{j=1}^{N} u_{j}  e^{-i2\upi k j/N},
\end{eqnarray}
\noindent
where $u_j$ is the velocity signal at $j$-th probe and $\hat{u}_k$ is the $k$-th Fourier mode.
Reorientations are classified using the amplitude, $A_1 =|\hat{u}_1|$ and phase, $\Phi_{1}= \tan^{-1} (\Imag~\hat{u}_1 / \Real~\hat{u}_1)$ of the first Fourier mode. During cessation-led reorientations, $A_1$ almost drops to zero, while in rotation-led ones the LSC rotates azimuthally without significant change in $A_1$ \citep{MishraDe}. Although these events are well explored experimentally in a cylindrical cell  \citep{Brown_Ahlers_JFM,Xi_Zhou_Xia_2006}, similar attempts in Cartesian box are rare.
In the following we capture the detailed dynamics of LSC in cubic box using the same methodology as followed in cylindrical geometries.


\par
\begin{figure}
\centerline{\includegraphics[width=8.cm]{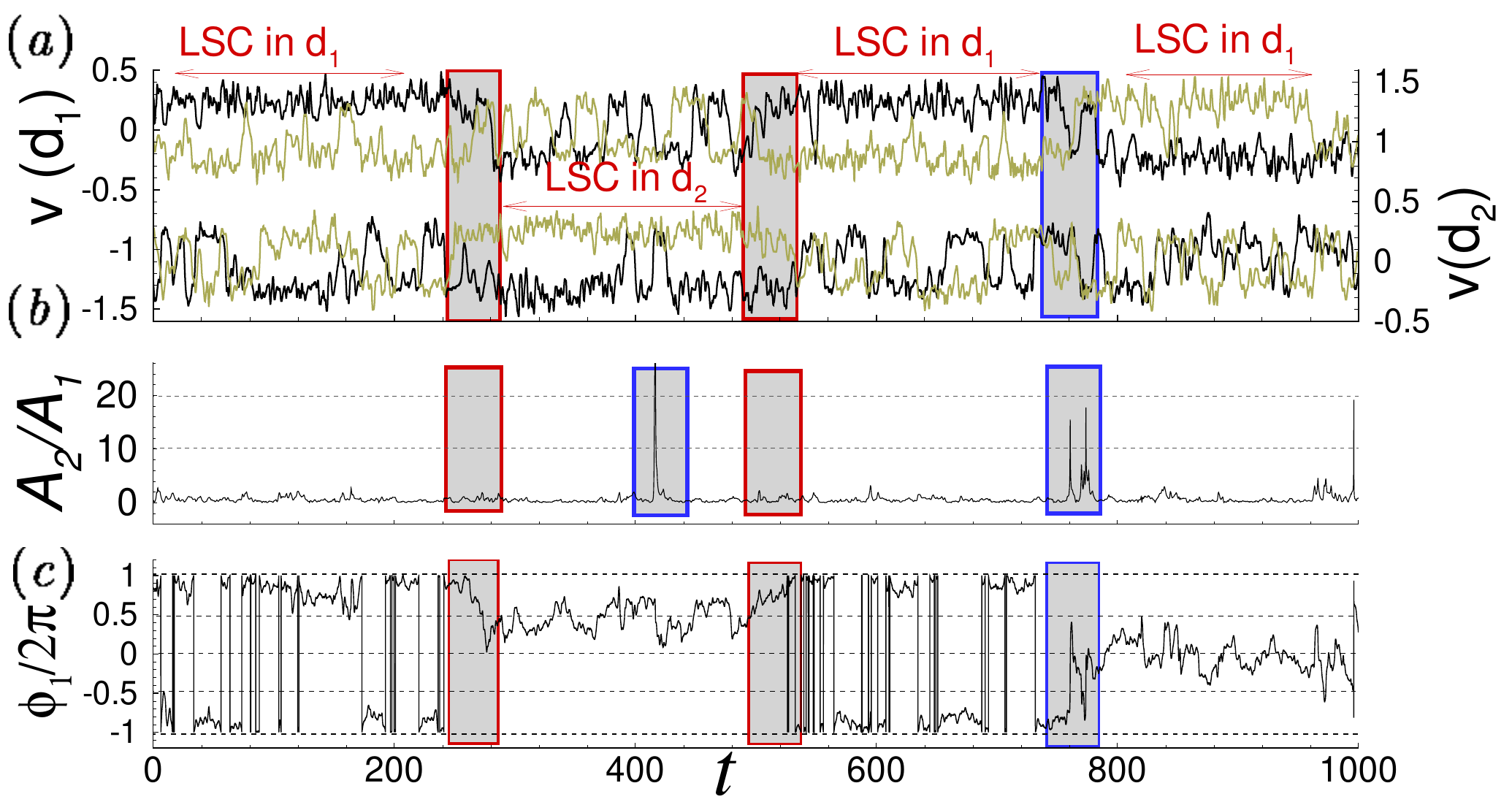} \includegraphics[width=5.35cm]{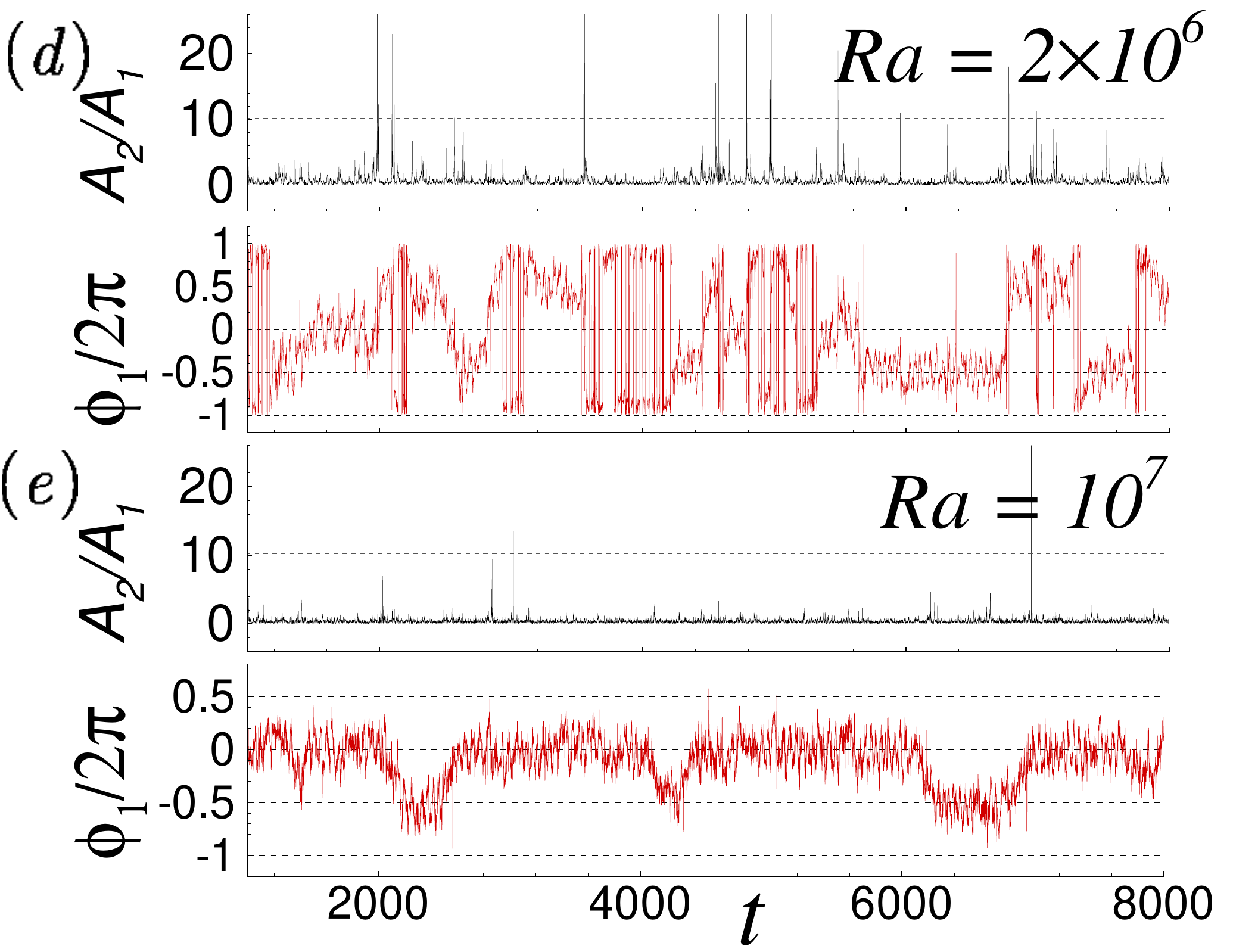}}
\caption{Time signals of (\textit{a}) vertical velocity along two diagonals, (\textit{b}) amplitude fraction $A_{2}/A_{1}$, and (\textit{c}) phase of the first Fourier mode $\Phi_1$ for $Ra =2\times10^6$. Complete time signature of $A_{2}/A_{1}$ and $\Phi_1$ for (\textit{d}) $Ra =2\times10^6$ and (\textit{e}) $Ra = 10^7$. }
\label{fig:timesig_reorientation}
\end{figure}

Time evolution of vertical velocity along the diagonals, amplitude ratio of the second and first Fourier mode ($A_{2}/A_{1}$), and phase of the first Fourier mode $\Phi_1$ are shown in figure  \ref{fig:timesig_reorientation}.
Initially LSC is observed along $\mathrm{d_1}$ where it stays for about $t\approx 230$ free-fall time units, following which it switches to $\mathrm{d_2}$ (see red windows) accompanied by a change in phase $\Delta \Phi_1 \approx \upi/2$ (figure \ref{fig:timesig_reorientation}\textit{c}) which indicates a partial reversal. $A_2/A_1$ remains negligibly small (figure \ref{fig:timesig_reorientation}\textit{b}) during this event, and thus it is identified as a rotation-led reorientation. A similar occurrence is observed at $t\approx 515$ where the LSC switches back from $\mathrm{d_2}$ to $\mathrm{d_1}$. However, at $t\approx 760$ ( blue windows), a different reorientation is observed as $A_2 / A_1$ shows a spike (figure \ref{fig:timesig_reorientation}\textit{b}) in addition to a phase change $\Delta \Phi_1 \approx \upi$. Here, the amplitude of the first Fourier mode drops considerably ($A_{1} \rightarrow 0$) and that of the second rises, and is identified as cessation-led reorientation. Note, though LSC persists along $\mathrm{d_1}$ before and after the reorientation, the sense of rotation reverses with a phase shift of $\Delta \Phi_1 \approx \upi$, resulting in a complete reversal. We have observed many such partial and complete reversals for $Ra =2\times10^6$ (figure  \ref{fig:timesig_reorientation}\textit{d}), while for higher $Ra$ reversals are restricted to partial ones as shown in figure \ref{fig:timesig_reorientation}(\textit{e}).


\par
Interestingly, at $t\approx 410$ a similar spike in the amplitude ratio is observed (figure \ref{fig:timesig_reorientation}\textit{b}), but the time signals show that LSC persists in $\mathrm{d_2}$ without any change in orientation. Here the LSC vanishes and reappears in the same diagonal with same sense of rotation resulting in no effective change in angular measure, $\Delta \Phi_1 \approx 0$. This is identified as a cessation, but not as a reorientation since there is no change in the orientation of LSC. In all previous studies of similar configuration, the reorientations were predominantly rotation-led \citep{Bai_etal_PRE_2016,Foroozani_etal_2017,Valencia_etal_IJHMT_2007}. It is interesting to note that, the cessation-led reorientations that are observed in cylindrical domains are rarely seen in cubic configuration. In the present study, we observe several cessations resulting in partial, complete or no reversal at $Ra=2\times10^6$, while for $Ra=10^7$ very few cessation events are observed (see figure \ref{fig:timesig_reorientation}\textit{d,e}).




\subsection{Mechanism of flow reversals}

So far, we find that the LSC is confined in one of the diagonal planes and it switches intermittently between them with time. The question arises: what are the mechanisms that lead to the change in flow topology in different planes. In this section, we analyse the flow during the partial and complete reversals of LSC.
 \begin{figure}
\centerline{\includegraphics[width=13cm]{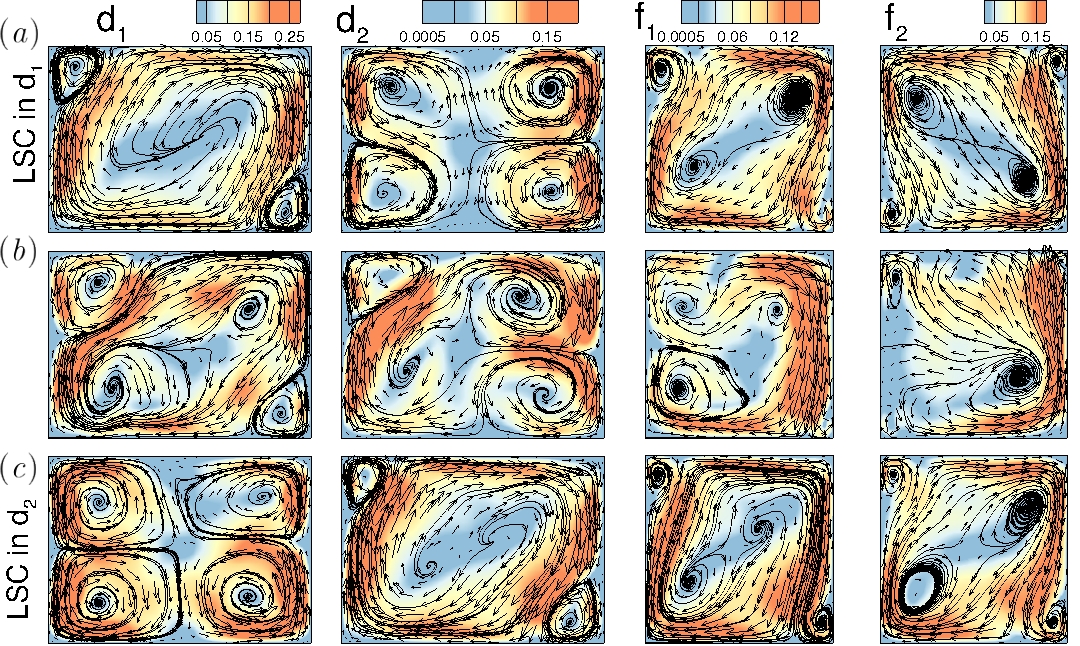}}
\caption{Velocity streamlines superimposed on the contours of absolute velocity at diagonal and face planes (\textit{a}) before, (\textit{b}) during, and (\textit{c}) after a partial reversal for $Ra = 2\times10^6$.}
\label{fig:Steamlines_d1d2}
\end{figure}

\par
The series of events that lead to a partial reversal for $Ra=2\times10^6$ is pictorially summarised in figure \ref{fig:Steamlines_d1d2} where time-averaged (10-20 convective time units for the transition and 150-200 for stable persistence) velocity streamlines are superimposed on contours of velocity magnitude. 
LSC is oriented along $\mathrm{d_1}$ (figure \ref{fig:Steamlines_d1d2}\textit{a}) as a clockwise primary roll accompanied by two counter-clockwise secondary corner-rolls, while in $\mathrm{d_2}$ a four-roll structure is noticed. 
During reversal, the corner-rolls in $\mathrm{d_1}$ grow in size (figure \ref{fig:Steamlines_d1d2}\textit{b}) while they induce the primary roll to break into two which results in a four-roll structure as shown in figure \ref{fig:Steamlines_d1d2}(\textit{c}).
However, in $\mathrm{d_2}$, two rolls with same sense of rotation grow and merge together to finally form the LSC.
Interestingly, flow topologies in both the diagonal planes are similar during the transition stage. 
In contrast to the diagonal planes, we observe a single-roll structure accompanied by two substructures inside the primary roll in the faces $\mathrm{f_1}$ and $\mathrm{f_2}$, which is similar to the abnormal single-roll state (ASRS) reported in the recent experimental study by \citet{Xin_etal_JFMR_2019}. Interestingly, after the reversal the sense of rotation of ASRS in one face reverses, while it remains same in the other.
In addition, we observe complete reversals in which the corner rolls in the LSC plane grow in size to yield a four-roll structure which further merge to produce an LSC with opposite sense of rotation. However, on the other diagonal plane the same four-roll structure is maintained during the entire reversal process. The sequence of flow structures and vortex reconnection in the LSC-plane during a complete reversal is very much similar to that observed in 2D/quasi-2D \citep{Sugiyama_Q2D_2010,chandra_Verma_2013} cases.
On the contrary, partial reversals are only realised in 3D geometry. 

\subsection{Energy transfer during reorientations}
Dynamical coupling between the activities in the diagonal planes is crucial to understand the reorientations of LSC. 
In order to address this aspect, we investigate the mechanism of energy transport by looking at the contributions coming from different terms (averaged over time and horizontal direction in $\mathrm{d_1}$ and $\mathrm{d_2}$ planes) that appear in the kinetic energy budget equation:
\begingroup\makeatletter\def\f@size{7}\check@mathfonts
\begin{equation}
\underbrace{\partial_t (\frac{1}{2}\overline{u^{\prime}_i u^{\prime}_i})+U_j \frac{\partial}{\partial x_j}(\frac{1}{2}\overline{u^{\prime}_i u^{\prime}_i})}_{A_k} =-\underbrace{\frac{\partial}{\partial x_j}\Big( \overline{u^{\prime}_j p^{\prime}}  -2\sqrt{\frac{Pr}{Ra}} \overline{u^{\prime}_i s^{\prime}_{ij}} +\frac{1}{2}\overline{u^{\prime}_i u^{\prime}_i u^{\prime}_j} \Big)}_{T_k}
-\underbrace{\overline{u^{\prime}_i u^{\prime}_j}S_{ij}}_{P_S}+\underbrace{ \overline{u^{\prime}_i\theta^{\prime}}\delta_i y}_{P_B}-\underbrace{2\sqrt{\frac{Pr}{Ra}}\overline{s^{\prime}_{ij}s^{\prime}_{ij}}}_{\bigepsilon}, 
\label{eq:tke}
\end{equation}
\endgroup
\noindent
where $u^{\prime}_i,~p^{\prime}$ and $s^{\prime}_{ij}$ are mean $(U_i,P,S_{ij})$ deducted fluctuations of velocity, pressure and strain rate, respectively. 
Here, $A_k$,$T_k$, $P_S$, $P_B$ and  $\bigepsilon$ represent the advection, transport, production by shear, production by buoyancy, and dissipation of turbulent kinetic energy (TKE), respectively.


\begin{figure}
\centering
\includegraphics[width=13cm]{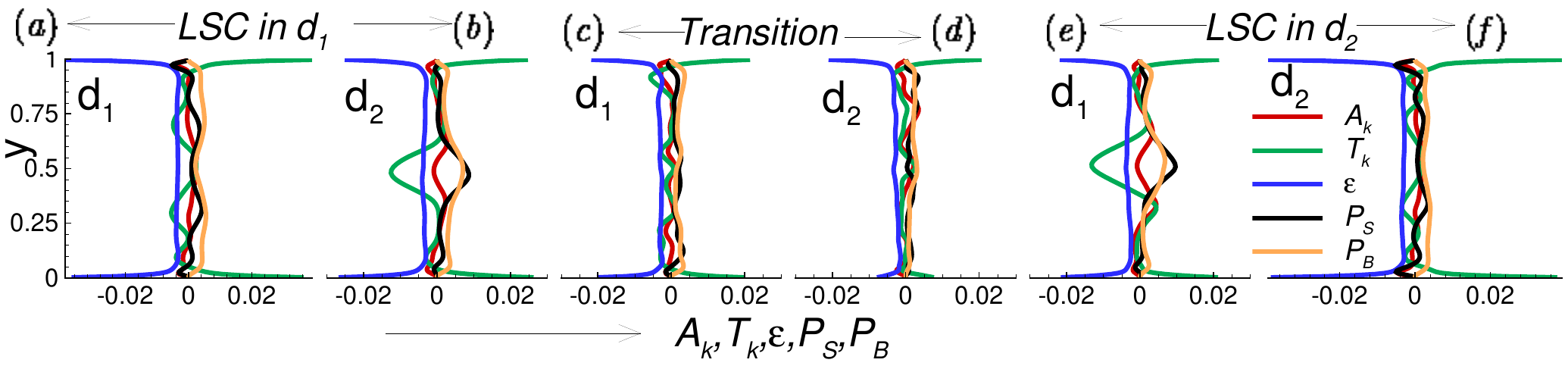}
\caption{Vertical profiles of TKE constituents in the diagonal planes, before (\textit{a-b}), during (\textit{c-d}) and after (\textit{e-f}) reversal.}
\label{fig:tkeAxial}
\end{figure}

\begin{table}
\begin{center}
\begin{tabular}{lccccccc}
Plane & $A_{k}(\times10^{-4}) $& $T_{k} (\times10^{-4})$ &$\bigepsilon (\times10^{-4})$&$P_{S} (\times10^{-4})$ &$P_{B} (\times10^{-4}) $&Imbalance$ (\times10^{-4}) $ \\ [3pt]
$\mathrm{d_1}$ (LSC in $\mathrm{d_1}$)&$2.48$&$-5.39$&$-42.9$&$12.8$&$39.1$&$8.99$\\
$\mathrm{d_2}$ (LSC in $\mathrm{d_1}$)&$3.75$&$\textcolor{red}{-13.4}$&$-39.7$&$21.9$&$35.9$&$\textcolor{red}{18.1}$  \\
$\mathrm{d_1}$ (Transient )           &$4.52$&$-7.33$&$-33.4$&$14.1$&$20.8$&$1.56$ \\
$\mathrm{d_2}$ (Transient )           &$0.33$&$-2.63$&$-28.4$&$14.7$&$17.5$&$3.71$ \\
$\mathrm{d_1}$ (LSC in $\mathrm{d_2}$)&$5.32$&$\textcolor{red}{-11.5}$&$-35.5$&$25.2$&$28.0$&$\textcolor{red}{17.7}$  \\
$\mathrm{d_2}$ (LSC in $\mathrm{d_2}$)&$3.16$&$5.57$&$-38.7$&$7.40$&$29.6$&$1.71$  \\
\end{tabular}  
\caption{Columns from left to right represent the average advection $A_k$, transport $T_k$, dissipation $\bigepsilon$, shear production ($P_s$) and buoyancy production ($P_b$) of TKE, along with the imbalance ($\bigepsilon + P_S + P_B$) in the diagonal planes $\mathrm{d_1}$ and $\mathrm{d_2}$.}
\label{tab:TKE_Budget}
\end{center}
\end{table}

The vertical variation of the different energy budget terms in $\mathrm{d_1}$ and $\mathrm{d_2}$ before, during, and after a reorientation is shown in figure \ref{fig:tkeAxial} . 
The buoyancy production remains small near the wall and increases as we move away from the wall to attain its maximum value. However, dissipation exhibits an opposite trend with maximum near the wall which asymptotically drops across the boundary layer.
Overall, the energy is mainly produced in the bulk, transported through large-scale structures and finally dissipated near the walls.
In the LSC-plane $\mathrm{d_1}$ (figure \ref{fig:tkeAxial}\textit{a}), shear and buoyancy production are significant at the interface of LSC and corner-rolls.
Unlike homogeneous shear flow turbulence, as fluctuations are largely sustained by buoyancy, a different mechanism than a simple $P_S \approx \bigepsilon$ is anticipated here. 
To make the energy transfer more quantitative we also compute the planar averaged quantities along with the imbalance between total production and dissipation as shown in table  \ref{tab:TKE_Budget}.
In $\mathrm{d_1}$, buoyancy production is dominant over shear (table \ref{tab:TKE_Budget}). However, in $\mathrm{d_2}$, both $P_B$ and $P_S$ are nearly comparable, and maximum production is observed in the bulk at the interface of the counter rotating rolls (figure \ref{fig:tkeAxial}\textit{b}).
Interestingly, shear production dominates over $P_B$ in the bulk unlike the LSC plane. 
Here, the imbalance (marked by red) is seen much higher than the LSC plane (18.1 over 8.99) which is mainly balanced by the transport term (negative peak in $T_k$ profile) which redistributes TKE in the domain. Thus, the excess TKE generated in the bulk is carried towards the walls, where it is finally dissipated.
During transient state, buoyancy and shear production are comparable which is balanced by dissipation rendering a trivial transport process (figure \ref{tab:TKE_Budget}\textit{c,d}). 
After reorientation, the LSC makes transition from $\mathrm{d_1}$ to $\mathrm{d_2}$ resulting in interchange of TKE profiles between these planes (figure \ref{fig:tkeAxial}\textit{e,f}). 
This leads to dominance of buoyancy in $\mathrm{d_2}$ where LSC is oriented and maximum imbalance in $\mathrm{d_1}$ (17.7 in table \ref{tab:TKE_Budget}) with a typical bulge in $T_k$ (see figure \ref{fig:tkeAxial}\textit{e}).

\begin{figure}
\centering
\includegraphics[width=10.5cm]{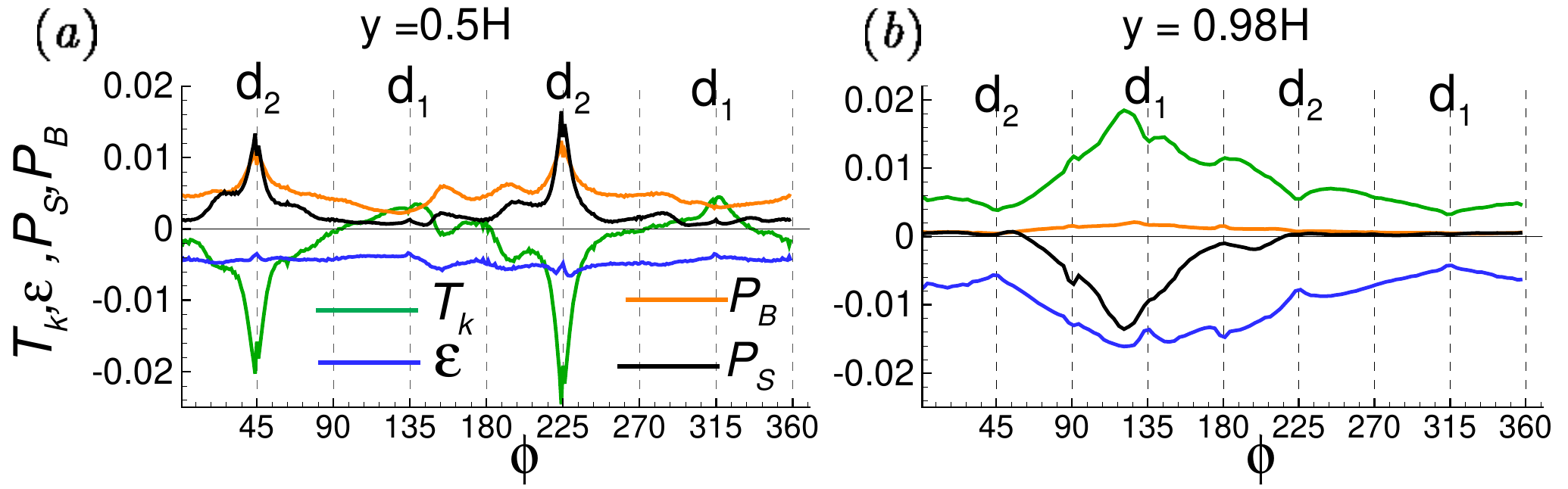}
\caption{Lateral variation of TKE constituents (\textit{a}) in the vertical mid-plane ($y=0.5H$) and (\textit{b}) inside the boundary layer ($y=0.98H$), when the LSC is oriented along $\mathrm{d_1}$. The dotted lines indicate the azimuthal locations of $\mathrm{d_1}$ and $\mathrm{d_2}$.}
\label{fig:tkeLateral}
\end{figure}

After having an idea about the energy distribution in the LSC and non-LSC planes, we are interested in identifying the mechanism which may be responsible for the reorientations. To this end we compute the azimuthal variation of different energy budget terms (equation \ref{eq:tke}) at two different vertical heights, namely, bulk ($y=0.5H$) and boundary layer ($y=0.98H$) for the situation when LSC is oriented along $\mathrm{d_1}$.
Both shear and buoyancy produce kinetic energy in $\mathrm{d_2}$ (figure \ref{fig:tkeLateral}\textit{a}) owing to a chaotic mixing which is extracted by the transport mechanism (note the -ve sign)
only to feed the mean flow in $\mathrm{d_1}$ (figure \ref{fig:tkeLateral}\textit{b}) where shear production is negative. Thus, a balancing act of shear between the LSC (mean) and high-variance (fluctuation) diagonals is evident which is fundamental to shear flow turbulence. Organised vertical motion of LSC results in a stronger correlation between the Reynolds stress and mean shear only in the boundary layer where horizontal motion becomes significant.
It is interesting to note that a major share of TKE is dissipated in the diagonal plane containing LSC, while its production occurs largely in the opposite plane, and the transport term redistributes TKE between these planes. Redistribution of TKE along the diagonals at different height, thus, plays a crucial role in sustaining the LSC as well as the coherent fluctuations in two opposite diagonals.


\par
\begin{figure}
 \centerline{\includegraphics[width=4.5cm]{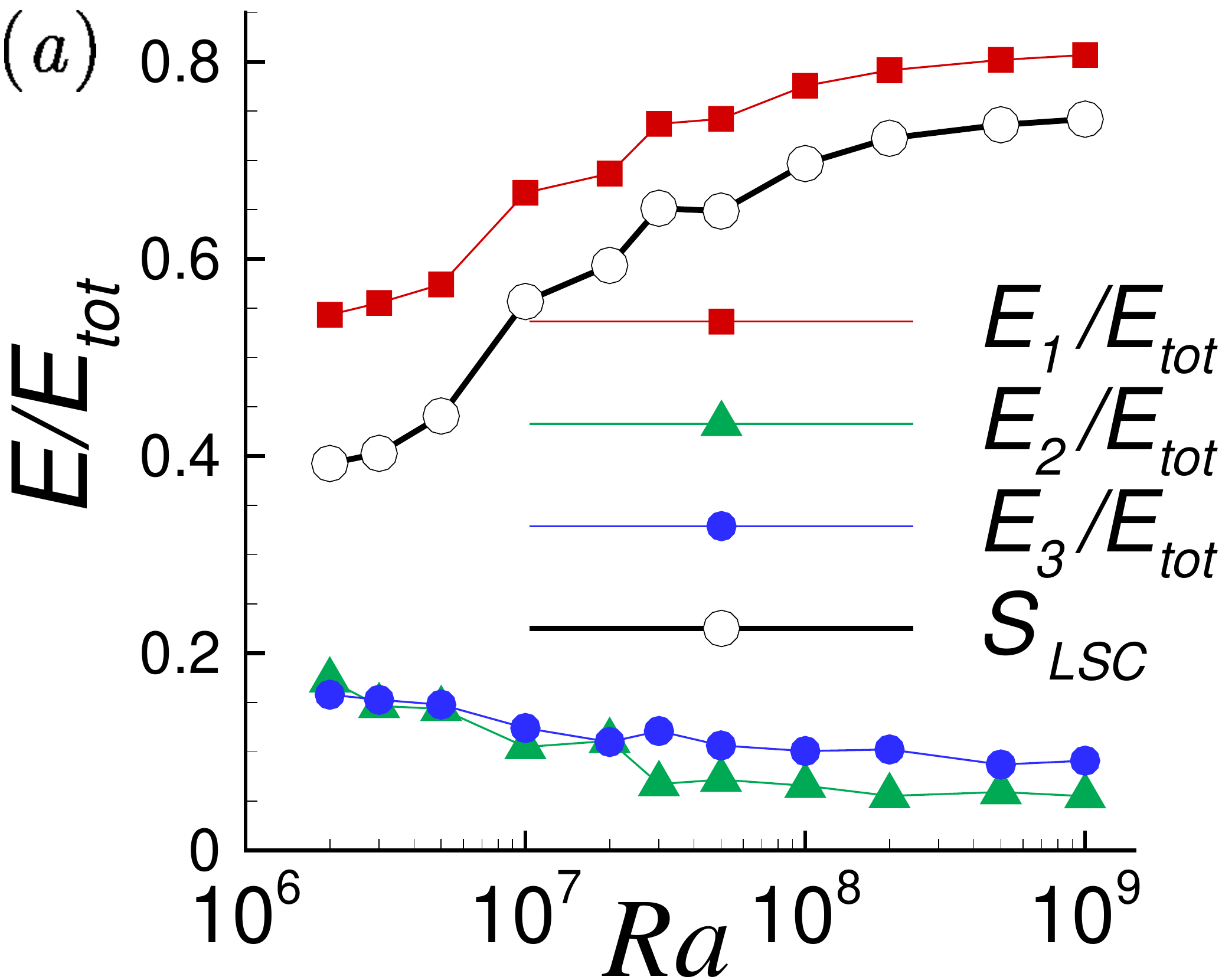}\includegraphics[width=8cm]{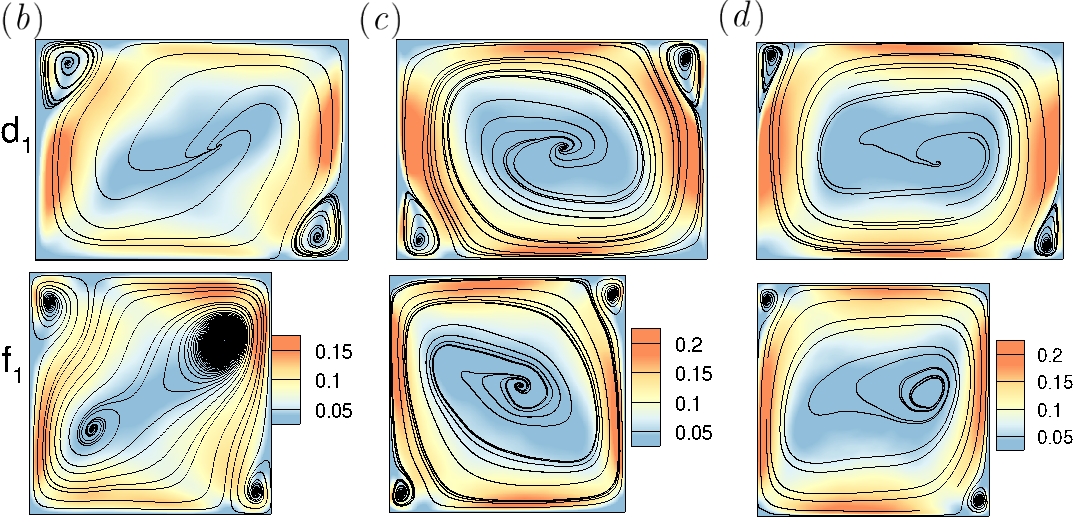}}
\caption{(\textit{a}) The variation of the energy fraction of the first thee Fourier modes ($E_{k}/E_{tot}$) and the strength of LSC ($S_{LSC}$) with $Ra$. Velocity streamlines superimposed on the contours of absolute velocity in the diagonal plane containing LSC and face $\mathrm{f_1}$ for (\textit{b})$Ra=2\times10^6$, (\textit{c})$Ra=10^8$, and (\textit{d}) $Ra=10^9$.}
\label{fig:EKSK_Steamlines_Ra}
\end{figure}

Next, we quantify the stability of LSC using the energy contained in the Fourier modes.
Figure \ref{fig:EKSK_Steamlines_Ra}(\textit{b}) shows the variation of energy fraction ($E_{k}/E_{tot}$) of first three Fourier modes with $Ra$, where $E_{k}=|\hat{u}_k|^2$ and $E_{tot}= \sum_{k=1}^{N/2} E_{k}$. Note that the energy content in the first mode increases steadily with $Ra$. At high $Ra$ ($Ra\approx 2\times10^8$) the first Fourier mode contains nearly $80\%$ of the energy indicating a clear dominance of LSC, which may be attributed to the decrease in the frequency of reorientations as observed in figure \ref{fig:timesig}(\textit{b,c}). In addition, the strength of LSC, $S_{LSC}=\mbox{Max} [(E_1/E_{tot}-2N^{-1})/(1-2N^{-1}),0]$ \citep{stevens7,Vishnu_PoF_2019}, is also shown alongside $E_{1}/E_{tot}$ which also follows the similar trend. 
Time-averaged velocity streamlines superimposed on velocity contours in the LSC-plane and face $\mathrm{f_1}$ for different $Ra$ are shown in figure \ref{fig:EKSK_Steamlines_Ra}(\textit{b-d}).
With increase in $Ra$ the LSC structure becomes more prominent and ``domain-filling" which effectively suppresses the corner-rolls. This observation is in line with the increase in strength of LSC with $Ra$. At higher $Ra$ the LSC becomes aligned with the horizontal direction causing nearly
unidirectional motion along its path. The ever diminishing four-roll structure on the other diagonal remains geometrically similar at higher $Ra$. As the corner-rolls play a pivotal role in flow reversals, decrease in the frequency of reversals at higher $Ra$ can thus be attributed to the suppression of these corner rolls.
Note the presence of substructures inside the single roll in the face plane for $Ra=2\times10^6$, which is absent at higher $Ra$. 
Recent study by \citet{Xin_etal_JFMR_2019} has shown that the emergence of substructures inside the LSC makes it less stable and thus increases the probability of reorientations. Hence, in addition to the corner-rolls, the structure and stability of LSC are also critical to its reorientations. With the increase in $Ra$ the substructures disappears, the LSC becomes more dominant and the corner-rolls get suppressed, and effectively the occurrence of reorientations reduces. 



\section{Conclusions}\label{sec:Conclusion}
Using a series of DNS calculations, dynamics of LSC in turbulent RBC inside a cubic box is investigated for $2\times10^6 \leq Ra \leq 10^9$.
LSC is primarily aligned along one of the diagonal planes accompanied by a four-roll structure in the other diagonal. 
For low $Ra$, the flow topology in the faces shows multiple substructures within the primary roll, which disappears at higher $Ra$.
The Fourier mode analysis of velocity signals in the mid-plane revealed the presence of rotation as well as cessation-led reorientations.
Further, we have analysed the flow topology in different planes for partial and complete reversals. 
The LSC weakens at the expense of growth of the corner-rolls to form a four-roll structure, while in the opposite plane two rolls merge together to form LSC.
Using kinetic energy budget, we have analysed the energy transfer mechanism that connects the dynamics between these planes.
We find that energy generated by the shear and buoyancy is transferred from the non-LSC plane to the LSC plane by the transport term.
With the increase in $Ra$, the LSC becomes more dominant and suppresses the corner-rolls and substructures. This effectively reduces the occurrence of reorientations at higher $Ra$.

\section*{Acknowledgement}

All simulations have been carried out in the `Param-Ishan' computing facility at IIT Guwahati.

\bibliographystyle{jfm}
\bibliography{reference}


\end{document}